\documentstyle[multicol,pre,aps,epsf]{revtex}
\begin{document}
\makeatletter
\def\endtable{%
\global\tableonfalse\global\outertabfalse
{\let\protect\relax\small\vskip2pt\@tablenotes\par}\xdef\@tablenotes{}%
\egroup\vskip 1pc
}%
\def\references{%
\list{\@biblabel{\arabic{enumiv}}}%
{\labelwidth\WidestRefLabelThusFar  \labelsep4pt %
\leftmargin\labelwidth %
\advance\leftmargin\labelsep %
\ifdim\baselinestretch pt>1 pt %
\parsep  4pt\relax %
\else %
\parsep  0pt\relax %
\fi
\itemsep\parsep %
\usecounter{enumiv}%
\let\p@enumiv\@empty
\def\theenumiv{\arabic{enumiv}}%
}%
\let\newblock\relax %
\sloppy\clubpenalty4000\widowpenalty4000
\sfcode`\.=1000\relax
\ifpreprintsty\else\small\fi
}
\makeatother
\draft

\title{
Search for universality in one-dimensional ballistic
annihilation kinetics.}
\author{Pierre-Antoine Rey and Michel Droz}
\address{D\'epartement de Physique Th\'eorique, Universit\'e de Gen\`eve,
CH-1211 Gen\`eve 4, Switzerland.}
\author{Jaros\l aw Piasecki}
\address{Institute of Theoretical Physics, Warsaw University, Ho\.za 69, 
Pl-00 681 Warsaw, Poland\vspace*{3mm}}
\author{\bf UGVA-DPT 1997/07-983\vspace*{6mm}}
\date{\today}
\maketitle

\begin{abstract}
We study the kinetics of ballistic annihilation for a one-dimensional
ideal gas with continuous velocity distribution. A dynamical scaling
theory for the long time behavior of the system is derived. Its
validity is supported by extensive numerical simulations for several
velocity distributions. This leads us to the conjecture that all the
continuous velocity distributions $\phi(v)$ which are symmetric,
regular and such that $\phi(0) \neq 0$ are attracted in
the long time regime towards the same Gaussian distribution and thus
belong to the same universality class. Moreover, it is found that the
particle density decays as $n(t) \sim t^{-\alpha}$, with $\alpha
\simeq 0.785 \pm 0.005$.
\end{abstract}

\begin{multicols}{2}
\section{Introduction}

Ballistically-controlled reactions provide simple examples of
non-equilibrium systems with complex kinetics and have recently
attracted a lot of interest \cite{EF,KS,BRL,R,jarek_uno,jarek_due,%
jarek_tre,jarek_four,jarek_five}. They consist of an assembly of
particles moving freely between collisions with given velocities. When
two particles meet, they instantaneously annihilate each other and
disappear from the system. In one dimension, it is enough to consider
point particles and we shall restrict ourselves to this case here. The
system with only two possible velocities $+c$ or $-c$ has been studied
in a pioneering work by Elskens and Frisch \cite{EF}. Using
combinatorial analysis, they showed that, in the long time limit, the
density of particles was decreasing according to a power law
$t^{-1/2}$ in the case of a symmetric initial velocity distribution;
Krug and Spohn \cite{KS} obtained independently similar results. Later
on, Redner {\it et al.}\ \cite{BRL,R} have studied the case of more
general velocity distributions. Based on numerical simulations and
mean-field type arguments they showed that the exponent characterizing
the power law decay of the particle density could depend on the
velocity distribution. The case of a general distribution has been
recently approached analytically by Piasecki \cite{jarek_uno} and Droz
{\it et al.}\ \cite{jarek_due,jarek_tre}. It was shown that the
annihilation dynamics reduced exactly to a single closed equation for
the two-particle conditional probability \cite{jarek_uno}. A method
which permits to solve this evolution equation for discrete velocity
distributions was developed in \cite{jarek_due} and explicitly applied
to the case of a symmetric three-velocity distribution. It turns out
that on one hand, different dynamical behaviors can occur depending on
the relative probability weights given to the three velocities and,
on the other hand, that the fluctuations are playing a very important
role, invalidating the predictions of mean-field or Boltzmann like
approaches.

The case of continuous velocity distributions was considered first by
Ben-Naim and co-workers \cite{BRL}. Within a Boltzmann approximation,
they showed that the particle density was decreasing according to a
power law $t^{-\alpha}$. Moreover, for initial velocity distributions
with a power-law dependence near the origin and a cut-off at $v_0$,
i.e.\ $\phi(v)\sim {|v|}^{\mu} \theta(v_0-|v|)$
($-1<\mu<0$), they found that the exponent $\alpha$ was non-universal
and was depending continuously on the value of $\mu$. This dependence
was confirmed by Monte-Carlo simulations. Note that the particular
case $\mu=0$ corresponds to a uniform distribution. In this case, and
in one dimension, numerical integration of the Boltzmann equation led
to $\alpha=0.77$, while Monte-Carlo simulations gave $\alpha=0.76$.
These values are very different from the ones obtained for the
discrete two- and three-velocity distributions.

The distributions considered by Redner {\it et al.}\/\ form a
very special subset of the continuous velocity distributions. So, it
would not be justified to generalize their conclusions. A deeper
understanding of the problem requires to study a wider class of
distributions. This was our motivation for this work. Several
questions could be asked. The main issue analyzed here, concerns the
existence of universality classes in this problem. Namely, does it
exist a class of continuous velocity distributions for which all its
members possess the same exponent $\alpha$, and what are the
characteristics of this class? As we have seen above, to approach this
question it is not sufficient to consider a mean-field like
approximation, but it is necessary to build up a theory taking into
account the fluctuations.

This paper is organized as follows. In Section \ref{model}, we define
the model and present the main ideas and conclusions obtained within the
analytical approach developed by Piasecki \cite{jarek_uno} and Droz
{\it et al.}\ \cite{jarek_due,jarek_tre}. Based on those results a
dynamical scaling theory is developed in Section \ref{scaling}, whose
validity is confirmed by extensive numerical simulations in
Section \ref{simulation}. Three different continuous velocity
distributions (Gaussian, uniform and Lorentzian) are studied. We found
that they are all attracted during the time evolution towards a
Gaussian distribution for which the decay exponent is $\alpha=0.785
\pm 0.005$. It is conjectured that all the continuous velocity
distributions regular near the origin and having a finite nonzero
value for $v=0$ are attracted by the same Gaussian distribution and
thus belong to the same universality class. Moreover, it is shown why
the velocity distributions with a power-law dependence with negative
exponent $\mu$ near the origin are not attracted by the Gaussian
distribution and can lead to non-universal exponents. Final remarks
are made in Section \ref{concluding} and the numerical algorithm used
in the simulations is explained in Appendix.

\section{The model} \label{model}

We assume that initially the particles are uniformly distributed in
space, according to the Poisson law, without any correlations between
their velocities. Note that other distributions than Poisson could be
considered as long as one is dealing with a renewal process
(see \cite{medi} for the definition of a renewal process). The
fundamental role in the analysis made in \cite{jarek_uno,jarek_due} is
played by the distribution of nearest neighbors. Suppose that at time
$t$ there is a particle at point $x_{1}$ in the fluid, moving with
velocity $v_{1}$. We denote by
\begin{equation}
\mu(x_{2},v_{2}|x_{1},v_{1};t) \label{1}
\end{equation}
the conditional probability density for finding its right nearest
neighbor at distance $x_{21} = x_{2} - x_{1} > 0$, with velocity
$v_{2}$. The density (\ref{1}) satisfies the normalization condition
\begin{equation}
\int d2\;\mu(2|1;t) = 1 \label{2}
\end{equation}
where a convenient short hand notation
\[
j \equiv (x_j,v_j), \qquad dj \equiv dx_jdv_j, \qquad j=1,2,\ldots
\]
has been used. As a rigorous consequence of the dynamics of ballistic
annihilation \cite{jarek_uno,jarek_due}, this quantity obeys the
equation:
\begin{eqnarray}
\lefteqn{
\biggl[ \frac{\partial}{\partial t}
      + v_1\frac{\partial}{\partial x_1}
      + v_2\frac{\partial}{\partial x_2} + C(1,2)
\biggr]\mu(2|1;t)}\hspace*{3mm} \nonumber \\
&=& \int\! d3\,[C(1,3)\mu(3|1;t) - C(2,3)\mu(3|2;t)]\mu(2|1;t) \nonumber \\
&&+ \int\! d3\!\int\! d4\,C(3,4)\mu(3|1;t)\mu(4|3;t)\mu(2|4;t). \label{eq:mu}
\end{eqnarray}
where $C(1,3)$ is the binary collision operator defined as:
\begin{equation}
C(1,2) = v_{12}\theta(v_{12})\delta(x_{21} - 0^+) \label{biop}
\end{equation}
$\theta(x)$ is the usual Heaviside function and $v_{12}$ stands for
$v_1-v_2$. If each particle has initially the same continuous
probability density $\phi(v)$ to move with velocity $v$, the state of
the system is translationally invariant. As a consequence,
$\mu(2|1;t)$ depends in the position space only on the distance
$x_{21}$ [we shall then write $\mu(x_{2},v_{2}|x_{1},v_{1};t) =
\mu(x_{21},v_{2}|0,v_{1};t)$]. A particular role is played by the
value of density $\mu$ at contact
\begin{equation}
\mu(0^+,v_2|0,v_1;t) = \lim_{x\to0\atop x>0} \mu(x,v_2|0,v_1;t) \label{cont}
\end{equation}
The notation $0^+$ stresses the fact that the distance between the
particles which are to collide approaches zero through positive
values. The higher order conditional distributions
$\mu_s(2,3,\ldots,s|1;t)$, $s=3,4,\ldots$ factorize for all times
$t>0$ as
\begin{equation}
\mu_s(2,3,\ldots,s|1;t)
 =  \prod_{j=2}^s \mu(j|j-1;t),\quad s=3,4,\ldots \label{facto}
\end{equation}
providing that this condition is fulfilled at time $t=0$. As a
consequence, $\mu(2|1;t)$ has the extraordinary property to obey
exactly the closed nonlinear integro-differential equation
(\ref{eq:mu}).

As we shall see later, it will be more convenient to consider
$\nu(x_2,v_2|x_1,v_1;t)\equiv\nu(2|1;t)$ the (conditional) density of
particles with velocity $v_2$, which are at a distance $x_2-x_1$ of a
given particle of velocity $v_1$. With Eq.\ (\ref{facto}) it follows
that $\nu(2|1;t)$ can be expressed as:
\end{multicols}
\vspace{-4.8mm}
\noindent\rule{20.5pc}{0.25pt}\rule{0.25pt}{5pt}
\begin{equation}
\nu(2|1;t)
 =  \mu(2|1;t)
 +  \sum_{s=3}^{\infty}\int\!d3\cdots\!\!\int\!ds\,
    \mu(2|s;t)\mu(s|s-1;t)\ldots\mu(3|1;t).
\end{equation}
We can then write a new closed equation governing the dynamics of
$\nu$ which, for a translationary invariant system, takes the form:
\begin{eqnarray}
\lefteqn{
\biggl[\frac{\partial}{\partial t} + v_{21}\frac{\partial}{\partial x}
      +v_{21} \theta(v_{12}) \delta(x)
\biggr]\nu(x,v_2|0,v_1;t)} \hspace*{2cm}\nonumber\\
&=& - \int dv_3
    \{\theta(v_{13})v_{13}\nu(0^+,v_3|0,v_1;t)
      [\nu(x,v_2|0,v_3;t)-\nu(x,v_2|0,v1;t)] \nonumber \\
& &   \hspace*{11mm}\hbox{}+\theta(v_{23})v_{23}\nu(0^+,v_3|0,v_2;t)
      \nu(x,v_2|0,v1;t) \nonumber \\
& &   \hspace*{11mm}\hbox{}+\theta(v_{32})v_{32}\nu(0^+,v_3|0,v_1;t)
      \nu(x,v_2|0,v3;t)
    \}. \label{eq:nu}
\end{eqnarray}
%\hspace*{22pc}\rule[-4.5pt]{0.25pt}{5pt}\rule{20.5pc}{0.25pt}%
%\vspace*{-8pt}

\begin{multicols}{2}
A complete description of the dynamic of the system is obtained once
the equation of motion for $f(v;t)$, the density of particles with a
velocity $v$, is given. It is readily obtained from the dynamics of
ballistic annihilation. One finds:
\begin{eqnarray}
\lefteqn{\frac{\partial}{\partial t}f(v_1;t)}\hspace*{8mm} \nonumber \\
&=& -\int dv_2\,\theta(v_{12})v_{12}\nu(0^+,v_2|0,v_1;t)f(v_1;t) \nonumber \\
& & -\int dv_2\,\theta(v_{21})v_{21}\nu(0^+,v_1|0,v_2;t)f(v_2;t).
\label{eq:evolf}
\end{eqnarray}

For a symmetric initial condition [i.e.\ $\phi(v)=\phi(-v)$], this
last equation takes a simplified form
\begin{eqnarray}
\frac{\partial}{\partial t}f(v_1;t)
&=& -f(v_1;t)\int dv_2\,
    [\theta(v_{12})v_{12}\nu(0^+,v_2|0,v_1;t) \nonumber \\
&&+\;\theta(v_{21})v_{21}\nu(0^+,-v_2|0,-v_1;t)], \label{eq:fv}
\end{eqnarray}
where we used that for a symmetric distribution
\[
f(v_1;t)\nu(x,v_2|0,v_1;t) = f(-v_2;t)\nu(x,-v_1|0,-v_2;t).
\]
To get a closed analytical form for the particle density $f(v;t)$, we
thus have to solve Eq.\ (\ref{eq:nu}), or at least to obtain the
density at contact $\nu(0^+,v_2|0,v_1;t)$. This is a very difficult
problem and despite numerous attempts, we have not yet been able to
complete this program. Accordingly, we choose a less ambitious
approach. In the next section, we shall propose a scaling theory for
$f(v;t)$ and $\nu(0^+,v_2|0,v_1;t)$ whose validity will be confirmed
by numerical simulations.

\section{Scaling theory} \label{scaling}

We are mainly interested in the behavior of the system in the long
time limit, where the time dependence of the physical quantities of
interest is given by power laws. Thus, it is natural to try to
develop for them a dynamical scaling theory. The validity of this
scaling theory will be attested {\it a posteriori}\/ by numerical
simulations.

For simplicity, we shall restrict our analysis to symmetric
distributions. For dimensional reasons, a time dependent
characteristic velocity should be introduced. For a symmetric
distribution, the average velocity vanishes for all times. However,
the mean-square velocity, denoted $M(t)$, does not vanish. Thus we
shall assume that ${[M(t)]}^{1/2}$ is the time dependent characteristic
velocity which will enter into the scaling theory.

Hence, we postulate the two following scaling laws:
\begin{eqnarray}
f(v;t) \label{eq:scaf}
&=& \frac{n(t)}{{[M(t)]}^{1/2}}\Psi\Bigl(v{[M(t)]}^{-1/2}\Bigr) \\
\lefteqn{\nu(0^+,v_2|0,v_1;t)
 =  \frac{n(t)}{{[M(t)]}^{1/2}}} \phantom{f(v;t)} \nonumber \\
&\times& \label{eq:scanu}
    \Phi\Bigl(0^+,v_2{[M(t)]}^{-1/2}\Big|0,v_1{[M(t)]}^{-1/2}\Bigr),
\end{eqnarray}
where $n(t)$ is the particle density at time $t$:
\begin{equation}
n(t) = \int\!dv\,f(v;t) 
\end{equation}
and
\begin{equation}
M(t) = \frac{\int\!dv\,v^2 f(v;t)}{\int\!dv\,f(v;t)}.
\end{equation}
It is useful to introduce the intermediate quantity [see Eq.\
(\ref{eq:evolf})]:
\begin{eqnarray}
I_{\rm c}(v_1;t)
&=& \int\!dv_2
    [\theta(v_{12})v_{12}\nu(0^+,v_2|0,v_1;t) \nonumber \\
&&\hspace*{6mm}+\;
     \theta(v_{21})v_{21}\nu(0^+,-v_2|0,-v_1;t)].
\end{eqnarray}
Using the scaling forms (\ref{eq:scaf}) and (\ref{eq:scanu}), one can
write:
\begin{eqnarray}
I_{\rm c}(v_1;t)
&=& n(t){[M(t)]}^{1/2}
    \biggl[\int\!du_2\,\theta(u_{12})u_{12}\Phi(0^+,u_2|0,u_1) \nonumber \\
&&+\;      \int\!du_2\,\theta(u_{21})u_{21}\Phi(0^+,-u_2|0,-u_1)
    \biggr] \nonumber \\
&\equiv&
    n(t){[M(t)]}^{1/2}J_{\rm c}(u_1),
\end{eqnarray}
where $u_i=v_i{[M(t)]}^{-1/2}$, ($i=1,2$). The time derivative of the
density and the mean square velocity can be written as:
\begin{eqnarray}
\frac{\dot{n}(t)}{n(t)} \nonumber
&=& -\frac{\int\!dv\,I_{\rm c}(v;t)f(v;t)}{\int\!dv\,f(v;t)} \\
&=& -n(t){[M(t)]}^{1/2}\frac{\int\!du\,J_{\rm c}(u)\Psi(u)}
                        {\int\!du\,\Psi(u)} \label{eq:dens}\\
\frac{\dot{M}(t)}{M(t)} + \frac{\dot{n}(t)}{n(t)} \nonumber
&=& -\frac{\int\!dv\,v^2I_{\rm c}(v;t)f(v;t)}{\int\!dv\,v^2f(v;t)} \\
&=& -n(t){[M(t)]}^{1/2}\frac{\int\!du\,u^2J_{\rm c}(u)\Psi(u)}
                        {\int\!du\,u^2\Psi(u)}. \label{eq:msv}
\end{eqnarray}
We note that {\sloppy $\int\!du\,J_{\rm c}(u)\Psi(u)/\!\!\int\!du\,\Psi(u)$
and $\int\!du\,u^2J_{\rm c}(u)$ $\times \Psi(u)/\!\!\int\!du\,u^2\Psi(u)$}
are two constants called $A$ and $B$ respectively. In the long time limit,
the solution of Eqs (\ref{eq:dens}) and (\ref{eq:msv}) is:
\begin{eqnarray}
n(t) &\sim& t^{-\alpha} \\
M(t) &\sim& t^{-2\beta}
\end{eqnarray}
with 
\begin{eqnarray}
\alpha &=& 2A/(A+B) \\
\beta  &=& (B-A)/(A+B).
\end{eqnarray}
For all values of $A$ and $B$, the scaling law:
\begin{eqnarray}
\alpha+\beta = 1 \label{eq:scaling_law}
\end{eqnarray}
holds. Note that this scaling law has already been obtained by
Ben-Naim {\it et al.}\ \cite{BRL} from heuristic arguments.

Using this scaling law, it is possible to justify {\it a posteriori}\/
our scaling postulate for $\nu$. Indeed, let us introduce the
dimensionless correlation function $g(x,v_2,v_1;t)$ satisfying
\begin{equation}
\nu(x,v_2|0,v_1;t) = g(x,v_2,v_1;t)f(v_2;t). \label{eq:g}
\end{equation}
\begin{figure}
\epsfxsize=6cm
{\samepage\columnwidth20.5pc
\centerline{\epsfbox[30 140 550 750]{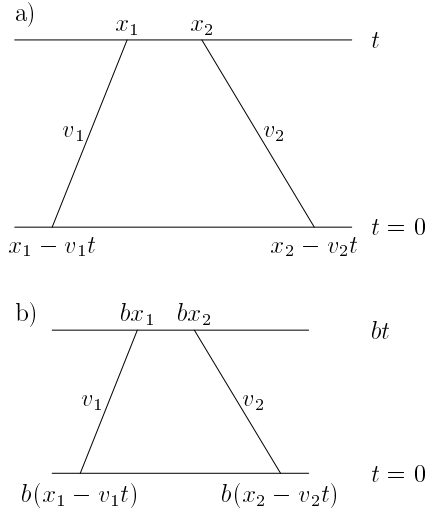}}
\caption{Illustration of the similarity relation satisfied by the
         correlation function $g(x_{21},v_2,v_1;t)$. a) A typical
         two-particle configuration is shown at time $t$. Initially the
         positions of the particles are $x_1-v_1t$ and
         $x_2-v_2t$. They move respectively with velocity $v_1$ and
         $v_2$. b Same configuration in which time and distances have
         been rescaled by a factor $b$ (the velocities remaining
         unchanged). The trapezoid defined by the points
         $b(x_1-v_1t)$, $b(x_2-v_2t)$ at $t=0$, and $bx_2$, $bx_1$ at
         $bt$ in Fig. b) is similar to the one defined by the points
         $x_1-v_1t$, $x_2-v_2t$ at $t=0$, and $x_2$, $x_1$ at $t$ in
         Fig a).\label{fig:sim}}
}
\end{figure}
%%%%%%%%%
As shown on Fig.\ \ref{fig:sim}, by rescaling both length and time by a
factor $b$, $g$ should satisfy the following similarity relation:
\[
g(x,v_2,v_1;t) = g(bx,v_2,v_1;bt).
\]
Moreover, rescaling the velocities by a factor $c$ corresponds to a
dilatation of the time scale by the same factor:
\[
g(x,cv_2,cv_1;t) = g(x,v_2,v_1;ct).
\]
Combining these two relations and putting $b=n(t)$ and $c={[M(t)]}^{1/2}$,
one immediately finds
\begin{eqnarray}
g(x,v_2,v_1;t)
&=& g\Bigl(n(t)x,v_2{[M(t)]}^{-1/2}, \nonumber \\
&&\phantom{g\Bigl(}
          v_1{[M(t)]}^{-1/2};n(t){[M(t)]}^{1/2}t\Bigr).
\end{eqnarray}
For sufficiently long time, the scaling law (\ref{eq:scaling_law})
implies that $n(t){[M(t)]}^{1/2}\sim t^{-1}$. Using Eq.\ (\ref{eq:g}) and
the scaling form (\ref{eq:scaf}) for $f$, one recovers then our
scaling postulate (\ref{eq:scanu}). This confirms the self-consistency
of our theory. Furthermore by introducing the scaling
postulates (\ref{eq:scaf},\ref{eq:scanu}) in (\ref{eq:fv}) one
obtains:
\begin{eqnarray}
\frac{\partial}{\partial t}f(v;t)
&=& -\biggl[\frac{3A-B}{2}\Psi(u) - \frac{B-A}{2}u\Psi'(u)
     \biggr] {[n(t)]}^2 \nonumber \\
&=& -{[n(t)]}^2 J_{\rm c}(u)\Psi(u),
\end{eqnarray}
hence an expression relating $\Psi(u)$ to $J_{\rm c}(u)$:
\begin{equation}
\frac{3A-B}{2} - \frac{B-A}{2}u\frac{\Psi'(u)}{\Psi(u)}
 =  J_{\rm c}(u). \label{eq:psiJc}
\end{equation}

Let us now suppose that, initially, $f(v;0) \sim
{|v|}^{\mu}\theta(v_0-|v|)$, with $-1<\mu<0$. In this case Ben-Naim {\it
et al.}\ \cite{BRL} argue that the long time behavior of $n(t)$ is
nonuniversal ($\mu$-dependent) and that $\Psi(u)$ retains the same
power-law dependence than $f$; in particular it diverges like
${|u|}^{\mu}$ for $u \to 0$. What could we say about this case within
our scaling approach? Letting $u\to0$ in Eq.\ (\ref{eq:psiJc}), we find
\begin{equation}
\frac{3A-B}{2} - \frac{B-A}{2}\mu = J_{\rm c}(0),
\end{equation}
or in terms of the exponent $\alpha$:
\begin{equation}
\frac{2\alpha-1}{1-\alpha} = \mu + \frac{2J_{\rm c}(0)}{B-A}.
\end{equation}
$J_{\rm c}(0)$ might in principle depend on $\mu$. However, it is very
unlikely that its dependence will be such as to exactly compensate the
linear term in $\mu$, hence the nonuniversality of the exponent
$\alpha$. On the contrary, if $\Psi(u)$ satisfies
\begin{equation}
\lim_{u\to0} u\frac{\Psi'(u)}{\Psi(u)} = 0 \label{eq:lim}
\end{equation}
[which is the case if $\Psi(u)$ and $\Psi'(u)$ are continuous functions
at $u=0$ and if $\Psi(0) \neq 0$], we find that
\begin{equation}
\frac{2\alpha-1}{1-\alpha} = \frac{2J_{\rm c}(0)}{B-A} \label{eq:exp_alp}.
\end{equation}
Hence $\alpha$ will be universal only if the right hand side does not
depend on the details of the initial distribution. We are
unfortunately unable to prove analytically this independence on the
initial conditions. However, in the next section we shall show that
this property is supported by precise numerical simulations for
several initial velocity distributions. Furthermore we shall find on
the same footing that $\Psi(u)$ is an universal Gaussian
function. Note that for $\Psi(u)$ to be universal, one should have
according to Eqs (\ref{eq:psiJc}) and (\ref{eq:exp_alp}) that $J_{\rm
c}(u)/(B-A)$ is universal.
 
\section{Numerical simulations} \label{simulation}

To test the validity of our scaling forms (\ref{eq:scaf}) and
(\ref{eq:scanu}), we have performed numerical simulations in one
dimension. The method we used is an exact synchronous time evolution
whose algorithm is detailed in Appendix. We considered the following
three different initial continuous velocity distributions:
\begin{itemize}
\item[i)] A Gaussian distribution:
\begin{equation}
\phi(v)
 =  \frac{1}{\sqrt{2\pi}v_0}\exp\biggl(-\frac{v^2}{2v_0^2}\biggr),
\end{equation}
with $v_0={[M(0)]}^{1/2}=1$.
\item[ii)] A uniform distribution with a cut-off:
\begin{equation}
\phi(v)= \frac{1}{2v_0}\,\theta(v_0-|v|),
\end{equation}
with $v_0=\sqrt3$, such that $M(0)=1$;
\item[iii)] A Lorentzian distribution
\begin{equation}
\phi(v)
 =  \frac{1}{\pi}\frac{v_0}{v_0^2+v^2},
\end{equation}
with characteristic velocity $v_0=1$.
\end{itemize}
For each simulation, we started with $2^{18}$ particles on a periodic
chain of length 2 \cite{bas_de_page}. Note that the initial
characteristic collision time $\tau_{\rm c}={[n(0)v_0]}^{-1}$ is of the
order ${10}^{-5}$. We choose to distribute the particles uniformly on the
line according to a Poisson law. However we verified that a regular
spacing distribution does not modify the long time dynamics of the
system.

\begin{figure}
{\samepage\columnwidth20.5pc
\epsfxsize=6cm
\centerline{\epsfbox{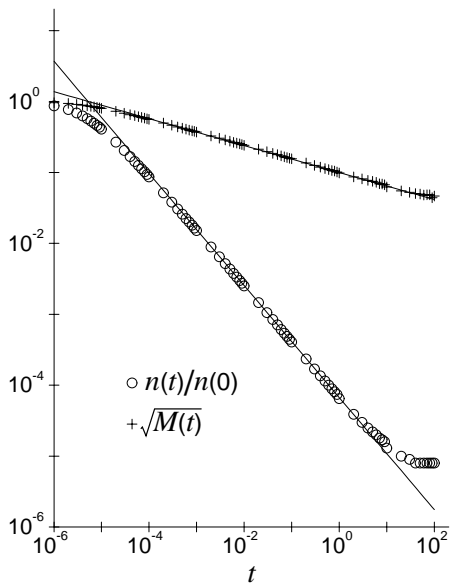}}
\caption{Plot of the relative density (circle) and the root mean square
         velocity (plus) as a function of time, for a Gaussian initial
         velocity distribution
         $\phi(v)=\exp(-v^2/2)/\protect\sqrt{2\pi}$. The periodic
         chain of length 2 initially contained $2^{18}$
         particles. Data are averaged over ${10}^4$ samples. The two
         straight lines are obtained by linear regression over a
         subset of points (typically for $t$ between ${10}^{-3}$ and 1).
         \label{fig:1}}
}
\end{figure}

\begin{figure}
{\samepage\columnwidth20.5pc
\epsfxsize=6cm
\centerline{\epsfbox{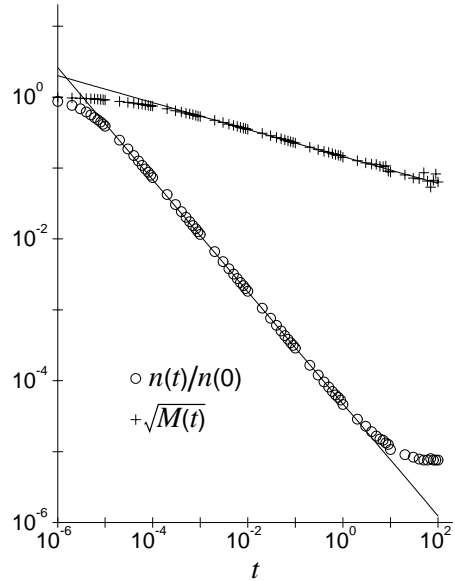}}
\caption{Plot of the relative density (circle) and the root mean square
         velocity (plus) as a function of time, for a uniform initial
         velocity distribution $\phi(v) =
         \theta(\protect\sqrt{3}-|v|)/\protect\sqrt{12}$, so that
         $M(0)=1$. The periodic chain of length 2 initially contained
         $2^{18}$ particles. Data are averaged over ${10}^3$
         samples. The two straight lines are obtained by linear
         regression over a subset of points (typically for $t$ between
         ${10}^{-3}$ and 1). \label{fig:2}}
}
\end{figure}

\begin{figure}
{\samepage\columnwidth20.5pc
\epsfxsize=6cm
\centerline{\epsfbox{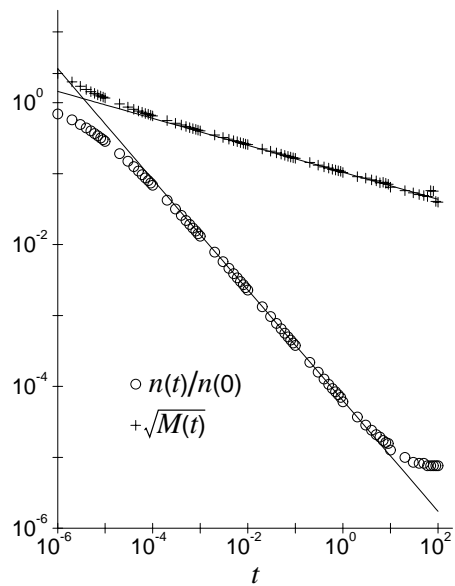}}
\caption{Same as in Fig.\ \protect\ref{fig:2} for a Lorentzian initial
         velocity distribution $\phi(v)=\pi^{-1}{(1+v^2)}^{-1}$.
         \label{fig:3}}
}
\end{figure}

The time dependent velocity distribution has been monitored yielding
the following results. First, we plotted on a double logarithmic scale
the number of particles (respectively the root mean square velocity of
a particle) as a function of time. The exponents $\alpha$
(respectively $\beta$) are then extracted from the slopes of the
curves. Linear regressions were made for various sets of points. The
retained value is the one corresponding to the best fit and the error
is given by the maxinum deviation. Typically, the fits were made for
time intervals such that ${10}^{-3} \lesssim t \lesssim 1$. For longer
times, the system starts to feel the boundary conditions. The lower
limit of the time interval corresponds approximately to the beginning
of the linear regime. Remark however that at this time about 98\% of
the particles have already disapeared. The results are given in Table
\ref{tab:1} and Figures \ref{fig:1}--\ref{fig:3}. Data were averaged
over ${10}^4$ samples for the Gaussian distribution and ${10}^3$ for
the uniform and Lorentzian cases, respectively.
%%%%%%%%%
\begin{table}%
{\columnwidth20.5pc
\caption{Numerical values of the exponents $\alpha$ and $\beta$
         obtained from the simulations (see
         Fig.\ \protect\ref{fig:1}--\protect\ref{fig:3}). The error is
         $\pm 0.005$ for each value of $\alpha$ and $\beta$. Results
         are given for three different initial velocity
         distributions.\label{tab:1}}
\begin{tabular}{cddd}
Velocity distribution & $\alpha$ & $\beta$ & $\alpha+\beta$ \\ \hline
Gaussian              & 0.785    & 0.195   & 0.980          \\
Uniform               & 0.795    & 0.195   & 0.990          \\
Lorentzian            & 0.780     & 0.195   & 0.975          \\
\end{tabular}
}%
\end{table}%
%%%%%%%%%
%%%%%%%%%
%%%%%%%%%

We see that the scaling law $\alpha + \beta = 1$ is well satisfied for
the different distributions. For the Lorentzian one however, the
results are less precise. This is due to the fact that the scaling
regime sets in for longer times, and longer simulations would be
needed to reach the same precision.

Note that for a uniform velocity distribution, Ben-Naim {\it et
al.}\ \cite{BRL} obtained $\alpha=0.76$ and $\beta=0.22$, but with a
slightly different algorithm (which involved diffusion in
addition to the ballistic motion) and a poorer statistics.

From the values quoted in Table \ref{tab:1}, we immediately deduce,
at least for the three different velocity distributions considered
here, that the exponents $\alpha$ and $\beta$ are the same. In view
of the arguments presented in Section \ref{scaling}, we are led to
conjecture that there is universality for a wide class of initial
conditions. However, this universality should not hold for initial
distributions which diverge or vanish at $v=0$ [see
Eq.\ (\ref{eq:lim})].

\begin{figure}
{\columnwidth20.5pc
\epsfxsize=6cm
\centerline{\epsfbox{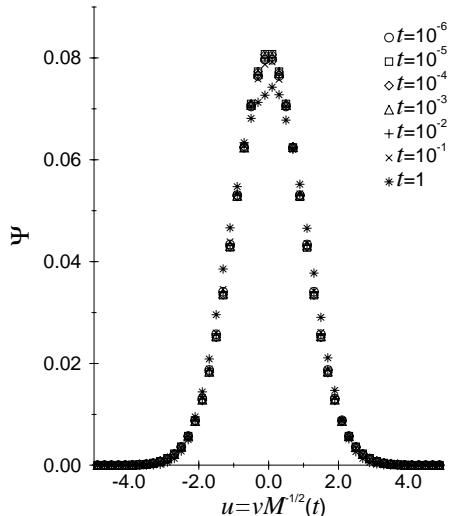}}
\caption{Plot of the scaling function $\Psi$ for a Gaussian initial
         velocity distribution (see Fig.\ \protect\ref{fig:1}) as a
         fonction of the reduced velocity $u=v{[M(t)]}^{-1/2}$ for 7
         different times. Data are averaged over ${10}^4$ samples. One
         remark that $\Psi$ keeps its Gaussian character as long as
         $t\leq 1$. \label{fig:4}}
}
\end{figure}

The scaling functions $\Psi(u)$ have also been measured for the three
different initial distributions. In Figs \ref{fig:4}--\ref{fig:6}, the
scaling functions $\Psi(u)$ are plotted versus the reduced velocity
$u=v{[M(t)]}^{-1/2}$. For the Gaussian case, one sees (Fig.\ \ref{fig:4})
that $\Psi$ keeps its Gaussian shape until $t\leq0.1$. In fact a finer
analysis of the six first even moments of $\Psi$ ($\langle
u^{2n}\rangle$ with $n=0,1,\ldots,5$) shows that $\Psi$ looses its
Gaussian character when $t\gtrsim 0.2$, i.e.\ when less than 50
particles remain in the system. Ultimately, when less than 10
particles remain in the system, $\Psi$ tends to a bimodal
distribution. We emphasize again that this late stage behavior is an
artifact of the finiteness of the system. In the thermodynamic limit,
the true asymptotic behavior would be Gaussian like. Similar
conclusions can be drawn for the uniform and Lorentzian cases (see
Figs \ref{fig:5} and \ref{fig:6}). Indeed, after a transient regime,
$\Psi$ adopts a Gaussian\hfill profile\hfill until\hfill
$t\gtrsim0.2$\hfill. These\hfill conclusions\hfill are
%%%%%%%%%
\noindent again confirmed by the analysis of the moments. Thus,
for these three distributions, we conclude that beside the universal
behavior of the exponent, the scaling functions $\Psi(u)$ are also the
same. There is an attractive Gaussian like scaling distribution in the
long time regime:
\begin{equation}
\Psi(u) = \frac{1}{\sqrt{2\pi}}\exp(-u^2/2).
\end{equation}
Here again, we are led to conjecture that such a behavior would be
valid for any velocity distribution which takes a finite non-zero
value at $v=0$ and which is regular near $v=0$.

\begin{figure}
{\columnwidth20.5pc
\epsfxsize=6cm
\centerline{\epsfbox{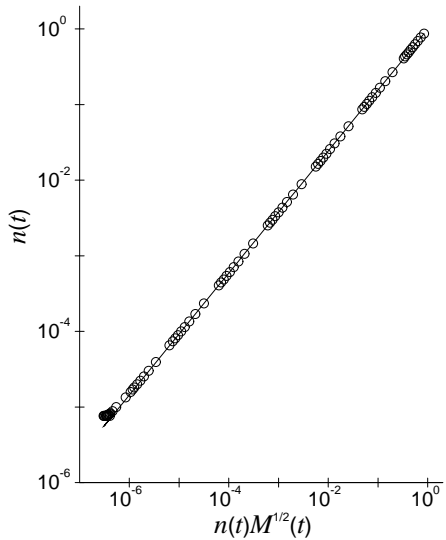}}
\caption{Log-log plot of the density as a function of $n(t){[M(t)]}^{1/2}$
         for a Gaussian initial velocity distribution (data are the
         same than Fig.\ \protect\ref{fig:1}, note that large values of
         $n(t){[M(t)]}^{1/2}$ correspond to short time). The comparison
         with a straight line is excellent both for short and long
         times. This shows that the range of validity of our scaling
         postulate is very wide for a Gaussian initial condition.
         \label{fig:7}}
}
\end{figure}

In addition, the range of validity of our scaling postulates
can be tested. Indeed from Eqs
(\ref{eq:dens},\ref{eq:msv}), one sees that
{\samepage
\begin{figure}
{\columnwidth20.5pc
\epsfxsize=6cm
\centerline{\epsfbox{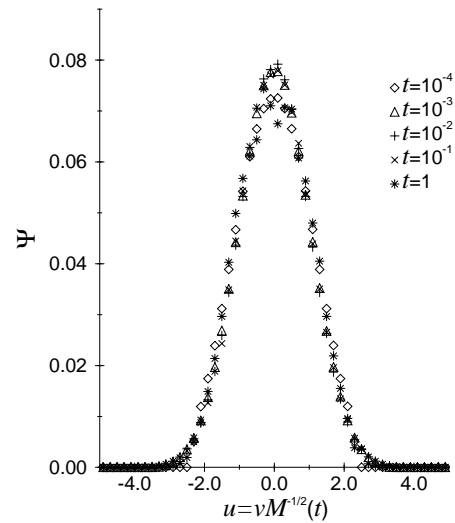}}\nopagebreak
\caption{Plot of the scaling function $\Psi$ for a uniform initial
         velocity distribution (see Fig.\ \protect\ref{fig:2}) as a
         fonction of the reduced velocity $u=v{[M(t)]}^{-1/2}$ for 5
         different times. Data are averaged over ${10}^3$ samples. One
         remarks that after a transient regime not represented here,
         $\Psi$ is attracted towards a Gaussian like scaling
         distribution (as soon as $t\protect\gtrsim {10}^{-4}$).
         \label{fig:5}}
}
\end{figure}}
%%%%%%%%%
{\samepage
\begin{figure}
{\columnwidth20.5pc
\epsfxsize=6cm
\centerline{\epsfbox{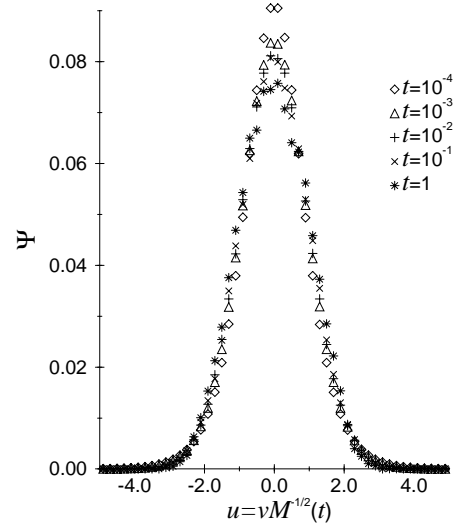}}
\caption{Same as in Fig.\ \protect\ref{fig:5} for a Lorentzian initial
         velocity distribution. \label{fig:6}}
}
\end{figure}}
%%%%%%%%%
\begin{equation}
n(t) = {\rm const}\ {\Bigl(n(t){[M(t)]}^{1/2}\Bigr)}^{\alpha},
\label{eq:nfM}
\end{equation}
so that the log-log plot of $n(t)$ versus $n(t){[M(t)]}^{1/2}$ should give
a straight line. In Fig.\ \ref{fig:7} we reproduce this plot for a
Gaussian initial distribution. We remark that Eq.\ (\ref{eq:nfM}) is
satisfied for times as short as $t\simeq{10}^{-6}$ (at this time less
that 20\% of the particles have already reacted) and as long as
$t\simeq5$. Note that the slope (i.e.\ the exponent $\alpha$) obtained
by linear regression is slightly larger than the value obtained from
Fig.\ \ref{fig:1}. This comes from the fact that the sum $\alpha +
\beta$ given by the value of the exponents measured on
Fig.\ \ref{fig:1} gives 0.980 rather than strictly 1. Multiplying the
slope of Fig.\ \ref{fig:7} by 0.980 reproduces indeed the value of
$\alpha$ quoted in Table \ref{tab:1}. For the uniform and Lorentzian
distributions the range of validity of the scaling postulates is
smaller, beginning only near $t\simeq{10}^{-3}$. This fact confirms the
particular role played by the Gaussian distribution in this problem.
%%%%%%%%%%%%%

\section{Concluding remarks} \label{concluding}
We have studied the kinetics of ballistic annihilation for a
one-dimensional ideal gas with continuous velocity
distribution. Starting from an exact analytical approach previously
derived, we established a scaling theory for the long time behavior of
such systems. The validity of this scaling theory has been tested
numerically for three different initial continuous velocity
distributions (Gaussian, uniform and Lorentzian). Both the dynamical
exponents and the scaling functions are the same for the three
cases. This led us to conjecture that all the continuous velocity
distributions which take a finite nonzero value at $v=0$ and which are
regular near $v=0$ are attracted towards a Gaussian distribution and
thus belong to the same universality class. Despite several attempts
we have not yet been able to prove this conjecture in the framework of
our exact analytical approach. We have only shown that a Gaussian
distribution is compatible, in the long time regime, with the exact
dynamical equation (\ref{eq:nu}). However, we have not been able to
prove that this Gaussian distribution was the only possible solution.

\section*{Acknowledgments}

Works partially supported by the Swiss National Science
Foundation. One of us (J.P.) acknowledged the hospitality at the
Department of Theoretical Physics of the University of Geneva were
part of this work was done, and the financial support by KBN
(Committee for Scientific Research, Poland) grant 2 P03 B 035 12.

{\appendix
\section*{Numerical algorithm}

To simulate the ballistic annihilation in one dimension, the simplest
algorithm is probably the standard molecular dynamics: starting from a
given configuration, one identifies the shortest collision time,
removes the two colliding particles and calculates the new positions
of the remaining particles at this time. Starting from this new
configuration, the process is iterated. This algorithm is very simple
but not very efficient, the computing time increasing with the number
of particles $N$ as ${\cal O}(N^2)$.

The numerical algorithm we used instead has been largely inspired by
the one developed by Krapivsky {\it et al.}\ \cite{R}. The idea is to
establish the list of all the (``true'') collision times arranged in
chronological order. From the initial condition we compute the
collision times of each particle with its right nearest neighbor and
sort those times into an ascending series, called $\cal A$, using a
standard sorting algorithm (see for example \cite{NR}). The shortest
time of this set corresponds to the first ``true'' collision time and
is the first member of the list of the ``true collision times'';
simultaneously it is removed from $\cal A$. Then one removes the pair
of particles $(n,n+1)$ corresponding to this first collision. As a
consequence, the collision times associated with the pairs $(n-1,n)$
and $(n+1,n+2)$ should be discarded from $\cal A$, producing a
truncated sorted list called ${\cal A}'$. The collision time of the
new nearest neighbor pair $(n-1,n+2)$ is computed and is the first
element of a new unsorted list, called $\cal N$. The process is then
iterated starting with the sorted list ${\cal A}'$ as long as its
first element is shorter than the shortest element of $\cal N$. When
this is no longer true, we merge both lists (${\cal A}'$ and $\cal N$)
into a new one which is sorted. This last list replace ${\cal A}$ and
the process continues until at most one particle remains in the
system. For the continuous velocity distribution we considered, this
merging step between ${\cal A}'$ and $\cal N$ occurs very rarely. For
example, for a Gaussian initial velocity distribution, it takes place
approximately hundred times for $2^{17}$ iterations of a system
containing initially $2^{18}$ particles, and the whole simulation used
about thirty seconds of CPU on a HP 9000 Serie 700 workstation. The
computing time increases with the number of particles $N$ roughly as
${\cal O}(N^{5/4}\ln N)$.
}
\end{multicols}
\centerline{\rule{25pc}{0.5pt}}
\begin{multicols}{2}

\end{multicols}
%%%%%%%%%%%%%

\begin{references}
\bibitem{bas_de_page}
        For open systems, the evolution does not notably differ as
        long as $t\lesssim2$, the time needed by a particle of
        velocity $v_0$ to totally cross the system.
\end{references}
\end{document}